\documentclass{article}
\usepackage{spconf,amsmath,amssymb,graphicx,booktabs}
\usepackage[T1]{fontenc}
\usepackage{times,microtype,balance}
\usepackage[hidelinks]{hyperref}
\usepackage{xcolor}
\definecolor{referenceorange}{RGB}{210,100,0}
\NewCommandCopy{\HCPoriginalcite}{\cite}
\RenewDocumentCommand{\cite}{o m}{%
  \IfNoValueTF{#1}{\textcolor{referenceorange}{\HCPoriginalcite{#2}}}%
    {\textcolor{referenceorange}{\HCPoriginalcite[#1]{#2}}}}
\let\originalthebibliography\thebibliography
\renewcommand{\thebibliography}[1]{\originalthebibliography{#1}\setlength{\itemsep}{-1.5pt}\setlength{\parsep}{0pt}\setlength{\parskip}{0pt}}
\newcommand{\HH}{\mathcal H}

\title{Reproducible Recurrent Patterns in Resting-State Brain Networks}
\name{Moo K. Chung
\thanks{Correspondence: mkchung@wisc.edu.}}
\address{University of Wisconsin, Madison, WI, USA}
\begin{document}
\ninept
\raggedbottom
\maketitle

\begin{abstract}
Spontaneous brain activity at rest exhibits structured relationships across regions and time. Whether its recurrent interactions carry a reproducible signature, rather than apparent recurrence arising from sampling and measurement noise, remains unclear. We investigate this question using repeated resting-state fMRI scans and network Hodge theory, which isolates circulation in directed functional networks. This algebraic representation places recurrent patterns in a common vector space, allowing them to be averaged and compared across time and individuals. The resulting patterns retain individual information across independent scans, while population means reveal a shared spatial organization. Together, these findings support reproducible recurrent organization at both individual and population levels.
\end{abstract}

\section{Introduction}
Resting-state functional magnetic resonance imaging (rs-fMRI) reveals organized interactions among brain regions even in the absence of an explicit task \cite{huang.2020.NM,finn.2015}. These interactions are commonly characterized through pairwise associations such as Pearson correlation \cite{huang.2020.NM}. Network models based on these associations describe how regional signals fluctuate together \cite{smith.2011}. Connectivity profiles can identify the same person across separate scans \cite{finn.2015} and reveal a common network organization shared across individuals \cite{gordon.2017,gratton.2018}. Symmetric association measures such as Pearson correlation do not distinguish the direction or ordering of interactions \cite{elyaagoubi.2025.TGI3}. Directed relationships have therefore been studied using Granger causality \cite{roebroeck.2005} and dynamic causal modeling (DCM) \cite{friston.2003,friston.2014}. Repeated-scan studies have reported reproducible features of estimated directed connectivity \cite{almgren.2018}, including whole-brain effective-connectivity estimates \cite{fraessle.2022}. Temporal sequences in resting-state activity \cite{mitra.2015} and recurring spatiotemporal patterns \cite{majeed.2011} further motivate examining how interactions act together across a network. Of particular interest are recurrent patterns, in which directed interactions form closed routes through several regions. Whether these patterns are reproducible and contribute individual information beyond conventional connectivity remains unresolved.

Comparing recurrent patterns across scans presents a methodological challenge. Individual connections have fixed anatomical labels, whereas recurrent routes can overlap, share connections, and change over time. Reproducibility therefore need {\it not} require identical cycles at corresponding moments in two scans; different sequences of activity may retain a similar overall organization. A suitable representation must capture this organization while allowing comparisons across scans and individuals. These comparisons must distinguish shared population structure from individual signatures: separate scans of the same person should be more similar than scans of different people.

In this study, we use network Hodge theory to separate directed flow into circulation and gradient components \cite{anand.2023.TMI,anand.2024.ISBI}. Representing circulation in a common vector space allows recurrent patterns to be averaged and compared without matching individual cycles \cite{chung.2026.ISBI}. Using repeated rs-fMRI scans from the Human Connectome Project (HCP) \cite{vanessen.2012}, we compute the temporal mean circulation within each scan and examine whether these summaries retain individual signatures alongside shared population organization. To test whether recurrence contributes information beyond conventional connectivity, we develop a contrastive connectivity fusion model that combines recurrent flow and Pearson connectivity through regional weights learned using a bidirectional softmax cross-entropy objective. We evaluate individual identification and assess the sensitivity of recurrent patterns to motion and physiological fluctuations.

\section{Methods}
\subsection{Repeated rs-fMRI scans}
We used HCP data \cite{vanessen.2012} from 445 participants with at least one eligible recording. The HCP resting-state protocol includes four scans acquired over two sessions, with left-to-right and right-to-left phase encoding in each session. Each scan contains 1,200 time points at a repetition time of 0.72 seconds, yielding 14.4 minutes of data. Recordings 1--4 denote REST1\_LR, REST1\_RL, REST2\_LR, and REST2\_RL, respectively.  The images underwent motion correction and spatial normalization through the HCP minimal preprocessing pipeline \cite{glasser.2013}, followed by ICA-FIX denoising \cite{salimikhorshidi.2014}. Signals were averaged within the 116 AAL regions spanning both hemispheres, subcortical structures, and the cerebellum \cite{tzourio.2002}. 

We then applied the frame-cleaning procedure in \cite{huang.2020.NM}: frames with framewise displacement above 0.5 mm, together with one preceding and two following frames, were excluded from fitting. Cosine-series reconstruction \cite{huang.2020.NM} interpolated these frames and applied the nominal 0.01--0.1 Hz bandpass. Recordings with more than 400 excluded frames were ineligible. No additional global-signal regression was applied. Directed interactions were estimated in 28-frame windows (20.16 seconds), advanced by one frame (0.72 seconds), yielding 1,173 windows per scan. This window length is comparable to the temporal extent of the hemodynamic response \cite{glover.1999}, providing a temporally localized summary of changing connectivity. Within each window, directed edge weights are
\begin{equation}
x_{ij}=\frac12\left[
\operatorname{Corr}(y_i(t),y_j(t+1))
-\operatorname{Corr}(y_j(t),y_i(t+1))
\right],
\label{eq:directed-association}
\end{equation}
where $y_i(t)$ is the signal in region $i$ at frame $t$. Correlations are computed within the window, with a one-frame lag of 0.72 seconds.

\begin{figure}[!t]
\color{black}
\centering
\includegraphics[width=\columnwidth]{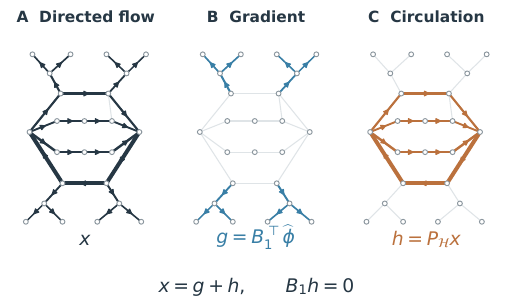}
\caption{\textbf{Extraction of harmonic flow.} In this schematic, directed flow $x$ (A) separates into its least-squares gradient $g=B_1^\top\widehat\phi$ (B) and harmonic circulation $h=x-g$ (C). Arrow widths indicate magnitude; gray edges carry zero flow. Removing the gradient leaves balanced inflow and outflow at every node.}
\label{fig:hodge-toy}
\end{figure}

\subsection{Recurrent interactions as diffusion limit}

Consider the dynamic complete graph $G=(V,E)$, whose node set $V$ consists of the 116 AAL regions. Within each window, we collect the entries $x_{ij}$ for $i<j$ into a signed edge-flow vector $x\in\mathbb R^{|E|}$, using $i\to j$ as the reference orientation. Let $B_1\in\mathbb R^{|V|\times|E|}$ be the node--edge incidence matrix, with $-1$ at the source node and $+1$ at the target node of each oriented edge, and zero elsewhere \cite{lim.2020}. The resulting 1-Hodge Laplacian defined over edges is $\Delta_1=B_1^\top B_1$ \cite{anand.2023.TMI}. We then diffuse noisy edge flow according to
\begin{equation}
\frac{dv(\tau)}{d\tau}=-\Delta_1v(\tau),\qquad v(0)=x.
\label{eq:hodge-diffusion}
\end{equation}
Its solution is $v(\tau)=e^{-\tau\Delta_1}x$. Unlike node diffusion under the graph Laplacian, which converges to a constant on a connected graph, edge-flow diffusion converges to a circulation in the cycle space \cite{anand.2024.ISBI}:
\[
\lim_{\tau\to\infty}v(\tau)\in\ker(\Delta_1) = \mathcal{H},
\]
which is the vector space spanned by cycles \cite{chung.2026.EMBC}. The diffusion limit $h=\lim_{\tau\to\infty}v(\tau)$ is called the harmonic flow and computed by removing the gradient of node potential $\phi$ via least squares (Fig. \textcolor{referenceorange}{\ref{fig:hodge-toy}}):
\begin{equation}
\widehat\phi=\arg\min_\phi\|x-B_1^\top\phi\|^2.
\end{equation}
Removing the fitted gradient gives
\begin{equation}
h=x-B_1^\top\widehat\phi=P_{\HH}x,\qquad
P_{\HH}=I-B_1^\top(B_1B_1^\top)^\dagger B_1,
\label{eq:cycle-projection}
\end{equation}
where $\dagger$ denotes the Moore--Penrose pseudoinverse. As an orthogonal projection, $P_{\HH}$ cannot amplify additive perturbations:
\[
\|P_{\HH}(x+\delta)-P_{\HH}x\|_2
=\|P_{\HH}\delta\|_2
\leq \|\delta\|_2.
\]
Thus, harmonic flow is stable to perturbations in the original connectivity, with gradient perturbations removed entirely. When variability lies in the gradient component, projection reduces that variability and supports temporal and population averaging.

\begin{figure*}[!t]
\color{black}
\centering
\includegraphics[width=\textwidth]{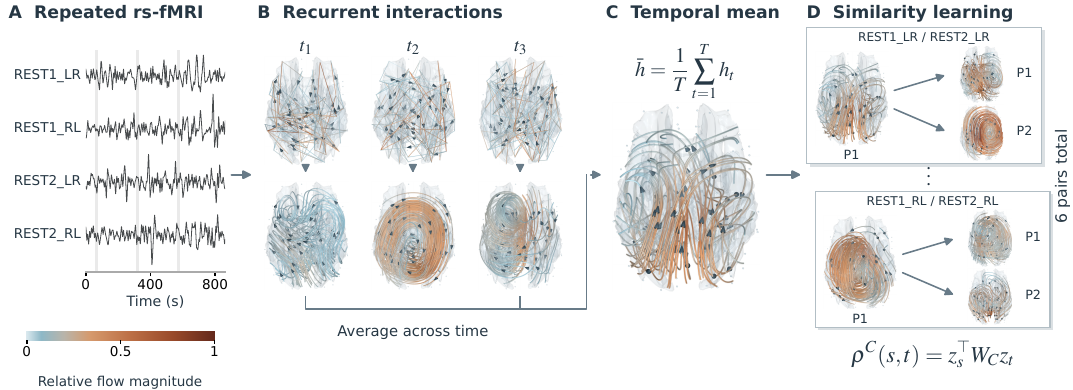}
\caption{\textbf{Workflow for learning cyclic interactions.}
\textbf{A.} Regional signals from the four HCP scans; shaded intervals mark the illustrated windows.
\textbf{B.} Recurrent organization at three time points, shown as directed graphs with overlapping loops (top) and streamlines interpolating the harmonic flow (bottom).
\textbf{C.} Temporal averaging gives the mean harmonic flow for each scan.
\textbf{D.} Two example recording pairs illustrate regional similarity learning across six pairs. In each panel, the upper arrow compares P1's mean flow across two scans, while the lower arrow compares P1's mean flow with P2's mean flow from the other scan.}
\label{fig:workflow}
\end{figure*}

Applying the above method to each window $t$, we obtain the harmonic flow $h_t$ and average it over time:
\begin{equation}
\bar h=\frac1T\sum_{t=1}^{T}h_t=P_{\HH}\bar x \in \HH,
\label{eq:recurrent-mean}
\end{equation}
where $\bar x=\frac1T\sum_{t=1}^{T}x_t$ is the temporal mean of the directed edge flows. The temporal mean $\bar h$ remains in the cycle space $\HH$ and therefore preserves circulation. Figure~\textcolor{referenceorange}{\ref{fig:workflow}} summarizes flow extraction, temporal averaging, and regional similarity learning.

We visualize harmonic flow using streamlines of a three-dimensional vector field $F$, obtained by Gaussian smoothing of signed edge vectors between anatomical region centers. The kernel standard deviation $\sigma$ was set to 1.5 times the median nearest-neighbor distance between region centers. These streamlines are integral curves ${\bf r}(\xi)$ satisfying \cite{mcloughlin.2010}: $\frac{d{\bf r}(\xi)}{d\xi}=F({\bf r}(\xi)),$ where $\xi$ parameterizes the curve.

\subsection{Regional weighting of cosine similarity}
Comparing recurrent interactions requires a similarity measure that preserves circulation direction. Existing approaches compare cyclic structure in undirected brain networks \cite{songdechakraiwut.2023} or establish correspondences between persistent cycles \cite{reani.2023}. These topological comparisons identify shared cycle structure \cite{garciaredondo.2024}, but do not distinguish opposite directions of flow around the same route. We therefore compare signed cyclic-flow patterns through regional weighting of cosine similarity  \cite{chen.2020,chiem.2022}, preserving directional information while allowing regional contributions to vary.

For scan $s$, let $\bar h_s$ denote the temporal mean harmonic flow. We subtract its sample mean and normalize it to unit length, denoting the resulting vector by $z_s$. We define the regionally weighted similarity between scans $s$ and $t$ as
\begin{equation}
\rho^C(s,t)=z_s^\top W_Cz_t,
\label{eq:regional-score}
\end{equation}
where $W_C$ is diagonal, with each edge weighted by the average of the learned weights of its two endpoint regions. Setting all weights to one gives ordinary cosine similarity. 

We apply the same similarity measure to Pearson connectivity using a separate regional weight matrix $W_P$, giving $\rho^P(s,t)$ and obtain the joint similarity:
$$
\rho(s,t)=\frac12\bigl[\rho^P(s,t)+\rho^C(s,t)\bigr].
$$
The regional weights for both components are learned jointly and shared across participants and scans.

\subsection{Learning and individual identification}

We included all 438 participants with at least two eligible scans, contributing 1,665 recordings. Five outer cross-validation folds evaluated identification accuracy, keeping all scans from each participant together, yielding 350--351 training participants and 87--88 evaluation participants per fold. The outer evaluation participants were excluded from fitting the centering reference, regional weights, and hyperparameters.

For a recording pair with $n$ training participants who had both scans available, let $s_i$ and $t_i$ denote the two scans of participant $i$, and define $\rho_{ij}=\rho(s_i,t_j)$. Thus, diagonal entries compare the same person across recordings, while off-diagonal entries compare different people. Although the scan similarity satisfies $\rho(s,t)=\rho(t,s)$, the entries $\rho_{ij}=\rho(s_i,t_j)$ and $\rho_{ji}=\rho(s_j,t_i)$ compare different pairs of scans and therefore need not be equal.

For fixed $\eta$ and $\lambda$, we learn the nonnegative regional weights defining $W_C$ and $W_P$ by minimizing the bidirectional softmax cross-entropy \cite{radford.2021} with a penalty that discourages departures from equal weighting:
\begin{equation}
\ell(W_C,W_P)
+\lambda\bigl[\|W_C-I\|_F^2+\|W_P-I\|_F^2\bigr],
\label{eq:loss}
\end{equation}
where
\[
\ell(W_C,W_P)\!\!
=-\frac{1}{2n}\sum_{i=1}^{n}\left[
\log\frac{e^{\rho_{ii}/\eta}}
{\sum_{j=1}^{n}e^{\rho_{ij}/\eta}}
+
\log\frac{e^{\rho_{ii}/\eta}}
{\sum_{j=1}^{n}e^{\rho_{ji}/\eta}}
\right].
\]
Here, $\lambda$ controls the penalty strength, and $\eta$ controls the sharpness of the softmax probabilities, with smaller values more strongly favoring the largest similarity \cite{radford.2021}. Each fraction gives the probability of the correct identity among $n$ candidates, and the two terms evaluate matching in opposite directions. Minimizing the loss encourages higher similarity between scans of the same person, while the penalty keeps edge weights near one. We pool the losses across all six recording pairs with equal weight per query and add the penalty in \eqref{eq:loss} once.

\section{Results}
\subsection{Individual identification}

\begin{table}[t]
\color{black}
\centering\fontsize{9}{11}\selectfont
\caption{Nested-cross-validation identification accuracy and 95\% confidence intervals, expressed as percentages. $\eta$ and $\lambda$ are the selected final-model parameters.}
\label{tab:identification}
\setlength{\tabcolsep}{2.5pt}
\renewcommand{\arraystretch}{1.15}
\begin{tabular}{@{}lrrrr@{}}
\toprule
Model & Accuracy & 95\% CI & $\eta$ & $\lambda$ \\
\midrule
Pearson & 71.55 & [69.09, 74.03] & 0.16 & $3.05\!\times\!10^{-7}$ \\
Directed & 37.20 & [34.76, 39.69] & 0.03 & $8.35\!\times\!10^{-5}$ \\
Recurrence & 57.27 & [54.70, 59.83] & 0.03 & $3.40\!\times\!10^{-5}$ \\
Pearson + Recurrence & 77.89 & [75.77, 80.05] & 0.48 & $9.65\!\times\!10^{-9}$ \\
\bottomrule
\end{tabular}
\end{table}

We evaluated 4,798 held-out queries from 438 participants with 1,665 eligible scans. Within each outer training set, three-fold participant-level cross-validation selected $\eta$ and $\lambda$ by maximizing identification accuracy through a best-first tree search over $10^{-4}\leq\eta\leq1$ and $0\leq\lambda\leq1$ \cite{varma.2006}. Confidence intervals used 10,000 participant-bootstrap resamples, keeping each person's repeated queries together.

Recurrence exceeded directed flow by 20.07 percentage points. Adding recurrence to Pearson connectivity improved accuracy by 6.34 points (Table~\textcolor{referenceorange}{\ref{tab:identification}}). All four models exceeded the 1.25\% chance reference. The results support the temporal mean harmonic flow as a reproducible individual signature that contributes information beyond Pearson connectivity.

\begin{figure*}[!t]
\color{black}
\centering
\includegraphics[width=\textwidth]{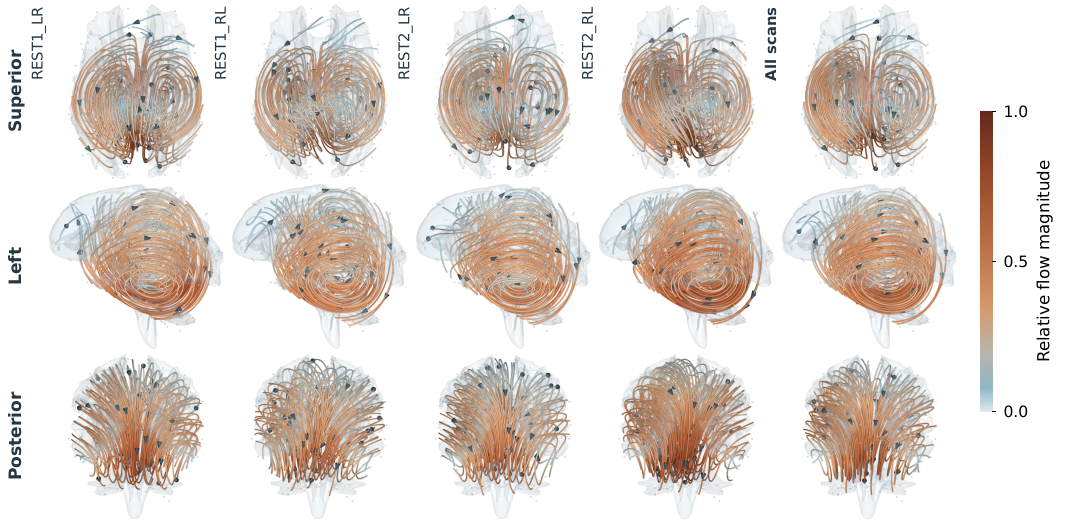}
\caption{\textbf{Population recurrent flow.} The first four columns show the recording-specific means; the last pools all 1,672 scans from 445 participants. We observed consistent cyclic interactions across recordings, revealing a reproducible population signature of recurrent flow.}
\label{fig:population}
\end{figure*}

\subsection{A shared population signature}
To examine shared organization, we averaged the temporal mean harmonic flows equally across eligible participants within each recording type (Fig.~\textcolor{referenceorange}{\ref{fig:population}}). We also pooled all 1,672 eligible scan means from 445 participants, with equal weight per scan. Each scan contributes 1,200 reconstructed time points and 1,173 overlapping windows: 2,006,400 time points and 1,961,256 harmonic-flow snapshots in total. The six pairwise cosine similarities between recording-specific means ranged from 0.953 to 0.971, indicating a highly consistent spatial organization across recordings. All six similarities exceeded chance expectations under random anatomical relabeling (Holm-adjusted $p=0.00060$) \cite{holm.1979}. This shared population pattern coexists with the individual differences used for identification.

\subsection{Sensitivity to motion and physiological fluctuations}
\begin{figure}[!t]
\centering
\includegraphics[width=\columnwidth]{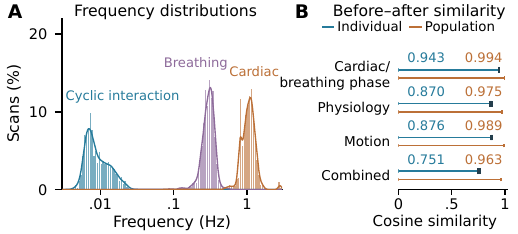}
\caption{\textbf{Sensitivity of recurrent flow to physiological and motion adjustment.}
\textbf{A.} Dominant-frequency distributions for circulation-strength fluctuations, breathing, and cardiac activity.
\textbf{B.} Cosine similarities between the original cyclic patterns and patterns recomputed after covariate adjustments.}
\label{fig:frequency-validation}
\end{figure}

We further investigated whether motion and periodic physiological fluctuations could contribute to the observed cyclic patterns. Dominant circulation-strength frequencies were estimated using Welch spectra with 256-sample segments \cite{welch.1967}. Across 1,672 scans, these frequencies ranged from 0.0041 to 0.0292 Hz. In the 1,469 scans with usable physiological recordings, dominant breathing frequencies ranged from 0.088 to 0.557 Hz (median 0.303 Hz), and cardiac frequencies ranged from 0.503 to 2.949 Hz (median 1.055 Hz). The dominant physiological rhythms were therefore faster than the observed circulation-strength fluctuations (Fig.~\textcolor{referenceorange}{\ref{fig:frequency-validation}}A).

To directly assess physiological effects, we regressed cardiac and respiratory phase terms from the regional signals \cite{glover.2000}, with additional adjustment for slower physiological fluctuations \cite{birn.2006,shmueli.2007} and motion \cite{power.2012}, and recomputed the harmonic flows. Participant patterns averaged their available matched scans across the four recordings; the population pattern pooled all 1,469 scans from 416 participants. After combined adjustment, the median individual cosine similarity was 0.751, while population similarity remained high at 0.963 (Fig.~\textcolor{referenceorange}{\ref{fig:frequency-validation}}B). Individual agreement exceeded chance under shuffled participant labels (Holm-adjusted $p\approx2\times10^{-6}$), and population agreement exceeded chance under random anatomical relabeling (Holm-adjusted $p\approx2\times10^{-6}$), using 1,000,000 permutations for each test. 
Thus, the shared population pattern and reproducible individual signatures are unlikely to arise  from the physiological and motion effects and may reflect intrinsic organization of human brain networks.

\section{Conclusion}
Harmonic flow captured reproducible individual recurrent interaction signatures and shared population organization across repeated rs-fMRI scans. Combining recurrence with Pearson connectivity improved identification, supporting its complementary value. Physiological and motion adjustment largely preserved the population pattern and retained individual correspondence.

\bibliographystyle{IEEEbib}
\bibliography{references}

\section{Compliance with Ethical Standards}
This study retrospectively analyzed existing HCP data. HCP participant recruitment and informed-consent procedures, including consent for data sharing, were approved by the Washington University institutional review board \cite{glasser.2016}. The University of Wisconsin--Madison Institutional Review Board approved this secondary analysis as exempt. No new participants were recruited or data collected for this analysis.

\noindent\textit{Acknowledgments.}
This work was supported by NIH MH133614 and NSF DMS-2010778. Data were provided by the Human Connectome Project WU--Minn Consortium (principal investigators David Van Essen and Kamil Ugurbil), funded through the NIH Blueprint for Neuroscience Research by 16 NIH Institutes and Centers (1U54MH091657) and by the McDonnell Center for Systems Neuroscience at Washington University.

\end{document}